# AI Conversational Tutors in Foreign Language Learning: A Mixed-Methods Evaluation Study

Nikolaos Avouris

avouris@upatras.gr
University of Patras, Patras, Greece

**Abstract**

This paper focuses on AI tutors in foreign language learning, a field of application of AI tutors with great development, especially during the last years, when great advances in natural language understanding and processing in real time, have been achieved. These tutors attempt to address needs for improving language skills (speaking, or communicative competence, understanding). In this paper, a mixed-methos empirical study on the use of different kinds of state-of-the-art AI tutors for language learning is reported. This study involves a user experience evaluation of typical such tools, with special focus in their conversation functionality and an evaluation of their quality, based on chat transcripts. This study can help establish criteria for assessing the quality of such systems and inform the design of future tools, including concerns about data privacy and secure handling of learner information.

**Keywords:** Foreign language learning, AI conversational tutors, Evaluation of interaction

## Introduction

The increasing demand for personalized and accessible language learning has led to the rise of AI-powered tutors. These tutors offer advantages such as their availability, independent of place and time, adaptive feedback, and interactive conversational practice, which are crucial for foreign language acquisition. In this context, while traditional learning methods (e.g., classroom instruction, human tutors) remain valuable, the AI tutors provide scalable and cost-effective alternatives, especially for speaking and conversation practice. This paper provides an overview of the field of AI conversational tutors in foreign language learning, focusing on chat-based and conversational modalities. It reports a user experience evaluation study and a comparative analysis of dialogues to assess quality dimensions. The objective is to highlight effective design patterns and discuss issues related to systematic evaluation of such tutors.

The research questions to be addressed are the following: RQ1: What are typical design patterns for AI conversational tutors in this domain? RQ2: How does the experience of interaction with AI tutors to general-purpose AI models compare? RQ3: What is a framework for measuring quality and evaluating such tutors and a method to automate the process?

This paper is structured as follows: First we review relevant literature on the state of the art. Then we report on a study, which includes three phases, (i) an expert-based evaluation of the conversational experience of using state-of-the-art AI tutors, (ii) comparison of the experience with that of interacting with general purpose AI tools, and then, (iii) discussion of an evaluation framework and an automated process for implementing it for measuring the quality of tutors. Finally, we discuss the implications of the findings of our study.

## Literature Review of AI language tutors

The integration of generative AI models into language learning platforms has revolutionized conversational practice, offering interactivity and personalization. Unlike earlier scripted chatbots, modern AI language tutors leverage large language models (LLMs) to simulate human-like dialogue, correct errors contextually, and adapt to learners' proficiency levels in real time. Conversational agents, in general, play an important role in supporting learning.

Yusuf et al. (2025) in their seminal paper, provide a conceptual framework of general applicability. With reference to this framework, language learning presents distinct characteristics, like the key role of natural language, need for real time error correction, ethical concerns due to voice data sensitivity, while the role of the human tutor is questioned within this context. Building on Labadze's et al. (2023) review, our study focuses specifically on AI conversational language tutors that simulate conversational practice and examines how they handle voice data—to effectively serve foreign-language learners.

In recent years, we observed the rise in the number of available tools for language learning with AI components. As an example, Duolingo (Shortt et al., 2023), the popular educational app, has introduced in 2023 an AI component to engage learners in conversation with an artificial character (Team, 2023). Early chatbots used predefined pathways and had many limitations, however in the post-2020 era, LLMs enabled open-ended conversations with contextual feedback. This accelerated new trend is based on natural language processing breakthroughs that allow for automatic speech recognition, as well as multimodality in interaction, where text and voice are often combined. For instance, Seo & Kim (2024) position LLMs as 'virtual conversation partners,' noting their ability to provide immersive, contextually adaptive practice across diverse topics. However, their research also cautions that such systems fail to correct subtle grammatical errors and often prioritize fluency over accuracy, potentially reinforcing learner mistakes. The issue of cultural authenticity in this context is important, as discussed by Godwin-Jones (2024), and Al-Othman (2024). In addition, Ye et al (2025) express their concerns about the danger of over-reliance on statistical patterns rather on communicative nuances. On the positive side, Lee (2025) underlines the multimodal, contextualized learning capabilities of such AI tutors, compared to isolated drills. An issue to take into consideration is the concern on ethical use of such tools. Neff et al. (2024) surveyed students and instructors of English as a foreign language and found out that learners worry about their recordings being used without their consent. Selvam & Vallejo (2025), study ethical and privacy concerns in the context of AI language tutors, highlighting the importance of fairness, linguistic diversity and human oversight. Despite the wide availability and use of AI tools for language learning, there seems to be limited number of empirical studies. Du & Daniel (2024) performed a systematic review on papers published in 2017-2023 on AI tutors and chatbots in language learning. They identify the advantages of speaking modality in chatbot interaction and the lack of empirical studies in this area. In addition, Zhang et al. (2024) provided a comprehensive survey of personalizing LLMs, highlighting the importance of personalization in interaction with LLMs and AI tutors. This is particularly important for foreign language learning, where the vocabulary, pronunciation, cultural background and linguistic level of the user should be taken in consideration in interaction with AI. Finally, other issues of design of the AI tutor, like the appearance of the tutor avatar, have been studied and are found to affect engagement, motivation, and sometimes learning outcomes (Tan et al. 2025).

In conclusion, the design of AI conversational tutors clearly involves many issues. To address this, our study focuses on developing a systematic approach to measure their quality. In the next section, we describe the first phase of our mixed-methods evaluation, which involved study of user interaction with typical AI tutors.

## User Experience with interacting with AI language tutors

Currently there are many applications for language learning, often in the form of games, past-time companions, or intelligent language learning systems. The focus of our study is on AI conversational tutors, so we established a method for selecting a representative set of them to perform our empirical study. Through information obtained from specialized blogs (e.g.

biglanguages.com, languagelearning thread in reddit), we selected for our study the following: *talkio, talkpal, univerbal, and langua*[1] *(coded as T1 to T4)*. All of them had received positive reviews for their state-of-the-art AI components. Typical screens of these applications are shown in Fig. 1.

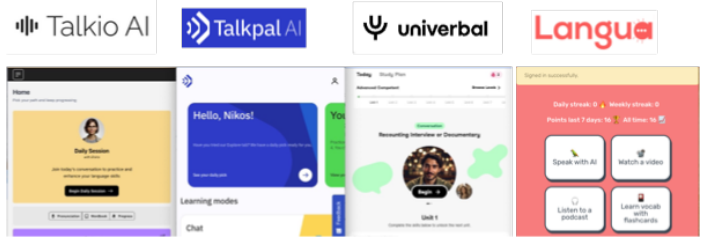

**Figure 1. The applications to study: T1 (*talkie*), T2 (*talkpal*), T3 (*univerbal*), T4 (*langua*)**

For each of the four systems, we followed the following steps. We created a learner profile (Spanish language learner, level B1/B2, preferred frequency of learning: 15' per day, personal interests: technology, sports, food). The four applications shared many common aspects: they were mainly designed for mobile use; they all established a daily and weekly plan, considering the personal interests of the learner. We spent time to familiarise with the applications and then we focused on interaction with the AI conversational tutor. We run a scenario of interaction which involved discussing a topic in a free chat for 5 to 10 minutes. Then we collected a transcript of the chat for the quality evaluation phase of this study. Next, we discuss the collected user experiences.

First, we report interaction with T1 (*talkio*). We personalized this application by selecting the tutor (Elvira), talking speed, and personal interests. In Fig.2 we can see typical views of the AI conversation tool. The dialogue can be represented as a sequence of chat messages, as in Fig 2(a), or alternatively see only the last turn of the tutor, or even no text. The modality of interaction can be either hands-free that do not necessitate pressing any buttons, or a manual mode that necessitates pressing the *Speak* button at the end of talk. This is a common design pattern, as we discuss next. The first design entails the risk of interpreting long pauses as end of talk, while the second is less natural and inhibits fluidity of interaction.

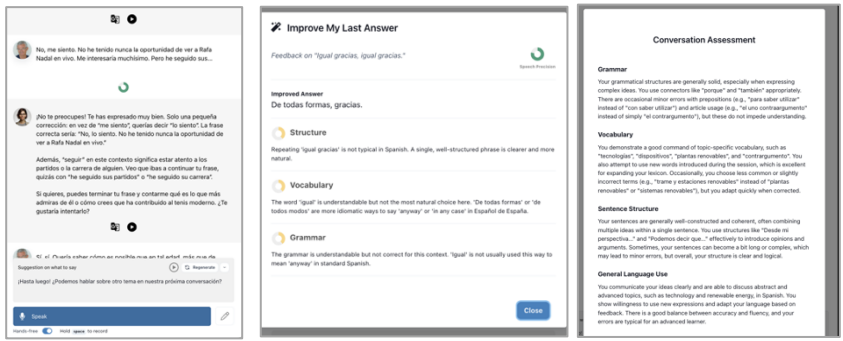

**Figure 2. Chat with T1 *talkio*, (a) interaction, (b) assess answer (c) final assessment**

---

[1] www.talkio.ai/app, app.talkpal.ai, chat.univerbal.app, languatalk.com/langua

There are options for presenting or not the response of the tutor, translate its response, or repeat the last phrase. There are also hints for the learner, (grey text area above the *Speak* button). The learner's utterances can be analysed according to a spoken precision index, see fig 2(b), that indicates how confidently the tutor has interpreted the learner's talk. The chat history can be exported and saved, while a final assessment of the chat is presented with comments and suggestions, see fig 2(c). This feedback, however, does not connect to the transcript of the chat.

The voice of the tutor was flat, with no emotion, and sometimes in hands-free mode the tutor interrupted the learner, when there is a longer-than-expected pause. However, it was observed that in case of unfinished utterance, the tutor tried to repair communication. Here is an example (translated from Spanish, extract from a chat about tennis): *Learner: No, I feel. I've never had the chance to see Rafa Nadal live. I'd be really interested. But I've followed his..."* (interruption due to timeout). *Tutor: I see you were about to continue your sentence, perhaps with 'I've followed his matches' or 'I've followed his career.' If you'd like, you can finish your sentence and tell me what you admire most about ... Would you like to give it a try?*

The flow of interaction was natural and the communication style supportive. The typical response of the Tutor contains an appraisal and encouragement, some remarks on errors, one suggestion on re-phrasing the argument, and finally a follow-up question. This makes the tutor utterance lengthier. Here is an extract of such a tutor message (translated from Spanish):

*"You have explained very well the two sides of social media use. [appraisal] Just a few small details to improve: the verb "apoyar" (to support) here is not the most appropriate ….[remarks on errors] I propose a more natural way to express it: 'For example, social media is a technology that helps us a lot …'.[suggestion for re-phrasing] Do you want to add any more arguments, perhaps related to the impact on concentration capacity or privacy? Do you think social media constitutes a real threat to mental health, or is it just one factor among many others? [follow-up]"*

Next, we proceeded with evaluation of user experience of interaction with T2 (*talkpal*). This application allows for many modalities of practice, like free chat/prescribed dialogues/ audio only conversation/ role play/ debates/ interaction with characters, etc. In Fig.3 we can see a typical view of the free chat with the application. The blue bubbles are the learner's talk. There are two modes of interaction, as with T1, one with automatic activation of the microphone and sending the message after a pause (it can be set between 1 to 5 secs), and a manual 'send message' mode. Dialogue representation is shown in Fig.3(a).

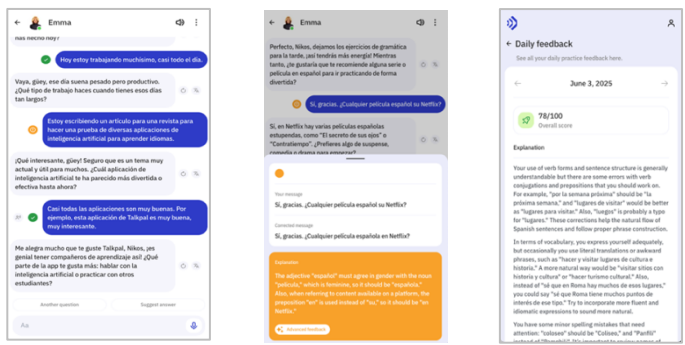

**Figure 3. Chat with T2, (a) chat window, (b) suggested correction, (c) daily feedback**

The avatar speaks empathetically and uses friendly tone; the messages are concise. Blue bubbles represent the learner's speech. A green sign next to a bubble means that the utterance is free of mistakes. If a yellow sign appears, an error is present; the learner can then select it for a suggested correction and explanation, see Fig. 3(b). This is a gamification element, as the user tries to make as few mistakes as possible, i.e. see fewer yellow signs. Breakdowns of communication are taken good care of. Here is an example: *Learner: Hola Emma, ¿podemos continuar nuestra charla? Quería hablar sobre… (timeout). Tutor: ¡Hola, Nikos! Claro, dime, ¿sobre qué te gustaría hablar hoy?*

There is no clear way to terminate the chat, if not by pressing the exit icon. Finally in the home screen there is an option for daily feedback with a summary of observations and suggestions and an overall score, as seen in Fig.3(c), with no reference to specific chat.

Next, we proceed with evaluation of T3 (*univerbal*). Compared to T1, T2, there were no user profile settings, beyond the areas of interest, T1 is stricter on executing the personal study plan, and seems more academic and less a casual language practicing app, as in the previous two cases. The interaction is not fluid. We tried to run the free chat using the '*Create your own conversation*' option, that leads to a menu for selecting the topic of the chat, and the person/place. After selecting various characters, it became obvious that this was not affecting the chat, for instance talking to a friend and to Yoel Harari about the 'evolution of technology' made no significant difference in the avatar picture and content of conversation.

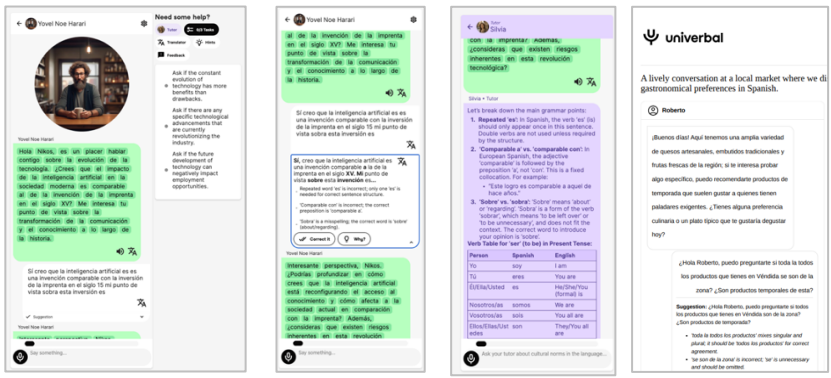

**Figure 4. Chat with T3, (a) chat window, (b) suggested correction, (c) Silvia's (the teacher) intervention to answer 'why', (d) annotated chat transcript**

The introductory utterance of T3 was longer than T2, with complex content. Learners were often cut off mid-sentence due to time limits, and no attempt was made to repair these communication breakdowns, see Fig.4(a). At the right panel, of fig 4(a), some hints are included. The learners' reply is tagged with the *Suggestion* option that opens a modal window, and as seen in fig 4(b) some corrections. If '*Why*' is selected, then the teacher comes in, fig 4(c), to provide some lengthy reference to the grammar. Interestingly, all words in the tutor's phrase were selectable for translation (green areas in Fig. 4), which ultimately hindered message readability. Finally, under '*Review*' the learner can find previous chats, and obtain an annotated transcript of the chat, see Fig 4(d).

Next, we discuss the user experience with T4 (*langua*). This application allows various modalities of speaking with AI: chat about anything/ role play/ debate/ games/ grammar practice. We have opted for the chat about anything. The voice of the selected character was particularly pleasant, and the flow of interaction and communication felt natural. Fig.5 shows typical views of the chat in this application. The chat settings included avatar selection, automatic vs manual turn taking, option to include or not a transcript, auto-translation, auto-correction, even selection of the Language Model to use, without any further explanation on how the latter would affect the user experience. In the default mode, each user message (in black in Fig.5) could be corrected, as seen in fig 5(b), repeated, translated, or alternative version suggested. With regards to the tutor messages, there were options for repeating the message, slow down, and translate it.

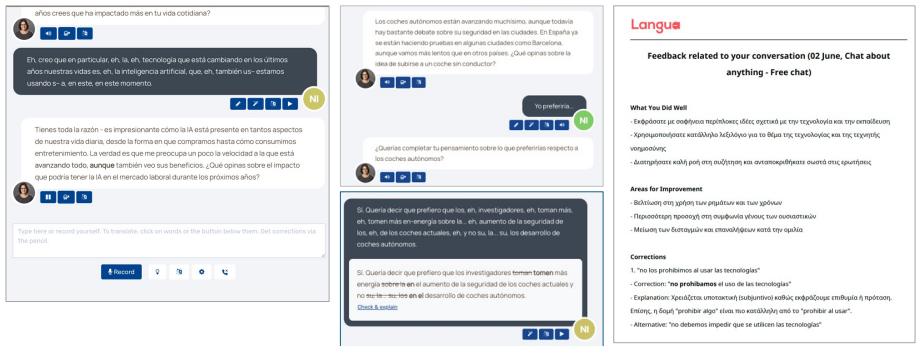

**Figure 5. Conversation with *T4*, (a) chat mode, (b) suggested correctiona and handling of unfinished phrase, (c) Feedback report**

The breakdowns of communication were taken care of, as seen in Fig 5(b) – upper part. In a case when the learner had not completed his/her phrase "*I would prefer…*" the response of the tutor was encouraging and contextual: '*Would you like to finish your thought about what you would prefer regarding autonomous cars?*'. The style of tutors' responses was in general supportive and constructive, without referring to mistakes. It focused on critical thinking over the learner's argument, trying to guide the learner into further reasoning. An example: *This is a very profound reflection – developing critical thinking and communication skills becomes even more crucial in this AI era. I find your point about not blindly accepting AI's answers but rather understanding the process behind them especially relevant. What methods do you suggest for cultivating that critical thinking in students when they interact with AI?*

In *T4*, like in the previous cases, the learner could go back to the chats, inspect them, play them back, continue the conversation, while a feedback document could be produced and downloaded, see Fig.5(c), that contained an annotated transcript of the chat and general comments and suggestions for improvement. No disclaimer or warning about storing or use of voice conversational data is included in the application.

Concluding this phase of our empirical study, we observe that the four applications share many common features. They allowed for free chat on topics selected by the user, supported alternative representations of the dialogue, often in typical chat mode, they provided contextual hints and suggestions to the learner, contextual corrections and link to the grammar for further explanations, suggested alternative phrasing of learner's message, they all provided a summary and feedback report. However, no consent was asked for storing private voice data. They also had distinct differences. The level of control allowed over the topic and length of learner's message was particularly low in *T3*, which made the conversation feel less natural. In addition, the same application did not acknowledge communication break downs, in case of learner's unfinished messages. The latency in tutor's response was a particular problem in *T1*, while the feedback was very informative in the cases of *T3* and *T4*, as they contained precise comments and annotated chat transcript. Finally, the style and content of tutor's utterances differed substantially, T1 (*talkio*) produced the lengthier messages, as it re-phrased the learner's message, the other tutors on average produced shorter messages. In Fig.6 we see the average words per turn for the four tutors.

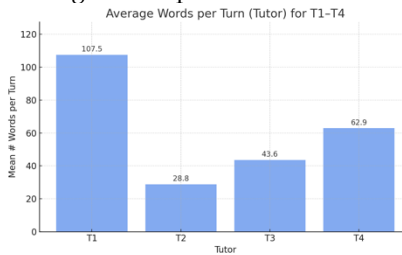

**Figure 6. Average words per turn for the four tutors**

## Interaction with general purpose AI conversational tools

To answer the second research question RQ2, we needed to study use of general-purpose AI conversational agents, in the role of an AI language tutor. The two available technologies at the time of the experiment that supported voice mode, were ChatGPT and Gemini. We used both and compared the experience with that of the tools described in the previous section. In Fig.7 we can see the starting state and the screen during voice interaction for Gemini and ChatGPT. The models used were *Gemini 2.5 flash* and *ChatGPT o4-mini*.

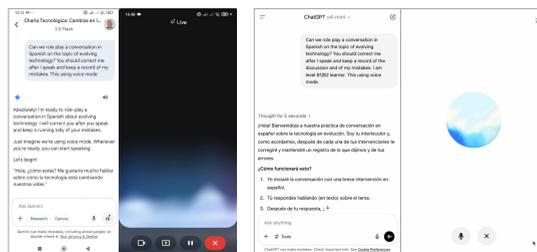

**Figure 7. Interaction with (a) Gemini and (b) ChatGPT**

The topic of discussion was in both cases that of evolution of technology. Here is the prompt used: *Can we role play a conversation in Spanish on the topic of evolving technology? You should correct me after I speak and keep a record of the discussion and of my mistakes. I am level B1/B2 learner. This using voice mode.* The interaction with both agents was fluid. Very few misunderstandings or broken communication occurred. Gemini decided to discuss the advances of AI, while ChatGPT opted for the change in use of mobile phones in the last 10 years. The total voice interaction time was for Gemini 6.47'' and for ChatGPT 10.45''. Gemini used a less empathetic voice and did not follow the instruction of providing feedback on mistakes of the learner, while ChatGPT did follow this pattern and thus was more verbose, so on average each turn of ChatGPT contained 130 words while that of Gemini, just 32. As a result, the learner was more verbose too in interacting with ChatGPT (average 61 words) compared with just 13 with Gemini. In this case there was no representation of the dialogue, quite different than the experience described in the previous section. This however did not seem to affect communication.

Finally, when the voice chat finished, the application returned in the usual text mode, where a transcript of the interaction was shown. However, we discovered that there were parts of the conversation missing from the script in both cases. Thus, to obtain a full record of the chat, we used a speech-to-text service, as this full transcript was needed for the next phase of the research. In conclusion, the experience in this case was comparable with that of the previous applications, however a serious limitation has been that of missing contextual corrections, as in the specialized AI language tutors, while a complete record, annotated for suggested improvement, was also missing.

## Quality evaluation of AI conversational tutors

In this final part of our evaluation study, we attempt to answer RQ3. We defined an evaluation framework and then applied it through an automated process to measure quality of the six tutors examined in the previous sections (coded T1 to T6, with T5=ChatGPT and T6=Gemini). It should be clearly stated that this part of the analysis is based purely on transcripts of typical chats and does not consider other aspects of user experience and functionalities of the tools, discussed in the previous sections. The evaluation framework was designed to assess tutor–learner conversations along ten complementary dimensions—eight focusing primarily on tutor and two on learner. Each dimension captures an aspect of communication or interaction quality in a foreign-language tutoring context. The evaluators had to score the Tutor on these ten dimensions, using a 0–5 scale (in 0.5 increments).

Our 10-dimensional framework draws inspiration from established practices in conversational agent evaluation, language pedagogy rubrics (e.g., CEFR), and prior work in dialog systems evaluation (e.g., Deriu et al., 2021; Liu et al., 2016). While adapted to the specific context of AI-based language tutoring, many of the dimensions—such as Value for Learning, Feedback, or Coherence—reflect criteria found in frameworks like CEFR, formative assessment rubrics in foreign language learning (Council of Europe, 2020), and dialogue evaluation metrics in NLP.

The dimensions of the framework were the following: *1. Value of learning*: The extent to which the tutor's contributions directly advance the learner's language acquisition related to provision of structured explanations, clear error corrections, and scaffolding questions that built on previous turns, *2. Supportive Style*: The tutor's friendliness, patience, encouragement, and overall warmth in guiding the learner, in the form of gentle corrections and positive re-enforcement. *3. Quality of Communication*: Clarity, grammatical accuracy, and appropriateness of the tutor's language at the learner's level. *4. Quality of Interaction:* How well the tutor maintained a coherent, responsive interaction with the learner, including turn-taking and adapting to learner input. *5. Coherence of Dialogue*: Logical progression and topic continuity in the conversation, for instance abrupt topic shift receive lower score. *6. Tutor Initiative*: The tutor's proactive leadership of the session—posing new questions, introducing ideas, or guiding toward learning goals, while tutors who respond passively score lower in this dimension. *7. Richness of vocabulary*: The variety and sophistication of the tutor's word choices—use of idiomatic expressions, domain-specific terminology, and advanced lexical items. *8. Value of Feedback:* The extent to which the tutor offers clear, accurate, and timely corrective feedback, e.g., explicit error correction, or reformulations.

In addition, the evaluators were asked to provide values for the dimensions: 9. *Learner Initiative* and 10. *Learner richness vocabulary*, to evaluate the quality of the activity, however without taking these two dimensions into account in the evaluation of the tutors.

To run the evaluation, we recruited five artificial experts (LLMs), among the most advanced ones, according to the https://lmarena.ai/leaderboard. We started with a superset and ended up with five that past a benchmark of calculating control values that had been before calculated by hand, i.e. count words per turn per tutor, or check agreement on quantitative values, like CEFR lexical evaluation of the interactions (Council of Europe, 2020).

The five artificial experts that were finally selected to participate in the study were the following: DeepSeek-V3, Grok 3, Gemini 2.5 Pro Preview 05-06, GPT-4-turbo, and Claude Sonnet 4. The prompt was the same; it described the framework for evaluation, provided them with anonymized chat transcripts and asked them to provide evaluation along the ten dimensions and justification of their judgement. Then we calculated the Inter-rater reliability (Cronbach's Alpha) for each one of the ten metrics, using the six tutors' scores. All metrics show at least "good" reliability (α ≈ 0.75–0.91). Value for Learning (α ≈ 0.897) and Value of Feedback Provided (α ≈ 0.911) exhibit the highest inter-rater consistency. The lowest—but still acceptable—reliability is for Quality of Communication (α ≈ 0.754). This is also shown in Fig.8 where the combined scores are shown. In this graph the 95% Confidence Intervals for most of the tutors are quite narrow, indication of agreement among the experts.

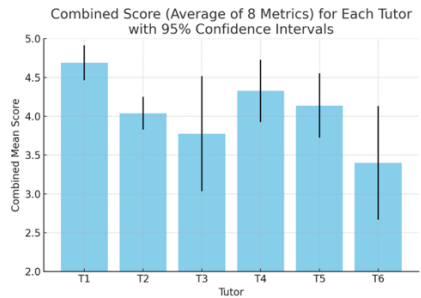

**Figure 8. Final evaluators' score of the six tutors**

From the overall evaluation report and the scores, we deduced that T1 and T4 achieved the highest overall scores on most metrics (especially on learning value, coherence, and interaction), while T5 also scored very highly on feedback, vocabulary, and supportive style. We should highlight that T3's role-play approach offered rich culture-specific vocabulary but delivered minimal explicit error correction, and T6 lacked clarity and timely feedback. It is interesting that the general use tutor ChatGPT (T5) scored higher than some of the more specialized tutors, judging from the chat transcript, not considering however the overall user experience, that was discussed in earlier sections.

## Conclusions and Future Work

This paper focused on studying AI conversational tutors for language learning. Using a mixed methods approach we evaluated the user experience, the main design patterns and assessed the quality of six different state-of-the-art tutors, including general purpose AI models.

Our study first involved identifying design patterns and limitations by simulating a typical use case. Following this, we employed artificial experts to evaluate the tutors' quality, drawing on transcripts of authentic conversations and guided by an ten-dimensional evaluation framework. From the study it is evident that the representation of the conversation, the contextual handling of learner errors and feedback without disrupting the flow of interaction is a key design requirement. From the evaluation of quality of the chat, it was found that issues like systematic corrective scaffolding, topic coherence, and rich vocabulary, that were identified as the main strengths of tutor T1, as well as the supportive style of T4 and the clear feedback of T5 should be considered as key quality targets in design of future AI tutors. On the other hand, the role-play approach of T3 raised questions on the effectiveness and its value. Finally, one interesting finding is that even widely available tools like T5, can produce high quality, rich conversational experiences.

We acknowledge several limitations in the reported study that might impact the broader applicability of our findings. Primarily, the qualitative empirical study involved only one expert evaluator and utilized a single chat transcript per tutor for the quality assessment. Our use of artificial experts in the quality evaluation phase could also be seen as a limitation. However, this aligns with contemporary trends in evaluation studies that employ artificial data and models. To address potential concerns, we pre-screened these artificial experts and involved more than one of them, measuring interrater reliability. The use of advanced language models as evaluators aligns with recent trends in AI-supported assessment (Atkinson & Palma, 2025), and our validation through inter-rater reliability strengthens confidence in their scoring. Nonetheless, we acknowledge that complementing automated evaluation with human expert assessment remains important and is planned for future work.

Ultimately, we hope that the proposed evaluation framework and its automated application will prove beneficial for future research into AI conversational agents.